\documentclass[12pt]{article}
\usepackage[margin=2.5cm]{geometry}
\usepackage{times}
\usepackage{graphicx}
\usepackage{amsmath,amssymb,bm}
\usepackage{authblk}
\usepackage{hyperref}
\usepackage{tikz}
\usepackage{comment}
\usetikzlibrary{arrows.meta,positioning,calc}

\title{\textbf{Programmable Non-Hermitian Synchronization of Light on a Silicon Photonic Processor}}

\author[1,*]{Ze-Sheng Xu}
\author[2,*]{Nan Cheng}
\author[3]{Mohammed S. Elmusrati}
\author[1]{Rohan Yadgirkar}
\author[1]{Andrea Cataldo}
\author[5]{Rui Wen}
\author[1]{Govind Krishna}
\author[4,*]{Jun Gao}
\author[1,*]{Ali W. Elshaari}
\affil[1]{Department of Applied Physics, KTH Royal Institute of Technology, Stockholm, Sweden}
\affil[2]{Department of Physics, University of Michigan, Ann Arbor, MI 48109-1040, USA}
\affil[3]{School of Technology and Innovations, University of Vaasa, 65200 Vaasa, Finland}
\affil[4]{School of Optical and 
Electronic Information, Huazhong University of Science and Technology, Wuhan 430074, China}
\affil[5]{Institute of Science Tokyo, Ookayama, Meguro-ku, Tokyo, 152-8550, Japan}
\affil[ ]{*Corresponding authors: zesheng@kth.se,\ nancheng@umich.edu,\ jungao@hust.edu.cn,\ elshaari@kth.se}

\date{}

\begin{document}
\maketitle

\begin{abstract}
\noindent
Synchronization is a pervasive collective phenomenon underlying the firing of neurons, the beating of the heart, and the coherent emission of lasers. 
Across these systems, dissipation plays an organizing role, suppressing microscopic differences and steering coupled units toward a common macroscopic order.
Here we harness engineered non-Hermitian dissipation to synchronize light directly in the optical domain.
Implementing non-Hermitian transition matrices on a silicon photonic processor, we drive arbitrary multimode optical fields toward a unique collective state with equal modal intensities and a globally locked phase — a process we call dissipation-induced phase synchronization.
The synchronization rate and total optical power throughput are independently programmable, enabling control over the dissipative dynamics without compromising reconfigurability.
These results recast dissipation as a functional resource and open a route to reconfigurable on-chip synchronization for classical and quantum photonic technologies.
\end{abstract}


\section*{Introduction}

Synchronization is one of the most widespread forms of collective order in nature. 
Since Huygens' observation in 1665 that two pendulum clocks suspended from the same beam could settle into a common rhythm~\cite{huygens1986pendulum}, synchronization has been identified in systems ranging from flashing fireflies and beating cardiac cells to neuronal rhythms associated with memory formation~\cite{pikovsky1985universal,fell2011role,strogatz2012sync}. 
In all of these examples, units with different microscopic states evolve toward a coherent macroscopic state. 
The remarkable feature of synchronization is that this collective order is not produced by eliminating interactions or losses, but through their combined action: coupling establishes collective degrees of freedom, while dissipation suppresses deviations from them.

The Kuramoto model provides the canonical theoretical picture of this mechanism~\cite{kuramoto1975international,acebron2005kuramoto}. 
A population of oscillators with different natural frequencies can spontaneously lock to a common phase once the coupling exceeds a critical threshold. 
This transition illustrates a broader physical principle: dissipation need not be merely destructive, but can erase microscopic differences and stabilize collective motion. 
This viewpoint has motivated extensive studies of synchronization in optical and quantum systems, including coupled lasers, optical oscillator networks, complex networks, and systems interacting through shared dissipative reservoirs~\cite{wu2007synchronization,uchida2012optical,ghosh2024recent,gupta2018statistical,gomez2007paths,arenas2008synchronization,skardal2014optimal}. 
It has also inspired dissipative approaches to quantum-state engineering, where carefully designed loss can generate and stabilize correlations rather than simply degrade them~\cite{diehl2008quantum,verstraete2009quantum,du2026decoherence}.

For coherent light fields, however, synchronization poses a distinct physical question. 
In closed coherent optics, linear evolution is unitary: it can redistribute amplitudes and phases among modes, but it cannot erase differences between arbitrary input fields or drive them toward a common state. 
Such convergence requires an open-system evolution in which some modal components are selectively attenuated relative to others. 
Non-Hermitian physics provides a natural framework for this mechanism. 
In open systems, gain and loss are encoded in the complex spectrum of an effective non-Hermitian Hamiltonian or evolution operator~\cite{ashida2020non,bergholtz2021exceptional,makris2008beam,ruter2010observation,peng2014parity,chen2017exceptional,feng2017non,pan2018photonic,miri2019exceptional,pyrialakos2025conservative}. 
By engineering this spectrum, one can make a desired collective mode dominant while dissipatively suppressing all others. 
Thus, synchronization of light can be viewed not as a threshold transition among nonlinear oscillators, but as a linear dissipative eigenmode-selection process.

In this work, we experimentally realize this idea by engineering non-Hermitian optical evolution on a programmable silicon photonic processor. 
We design and implement a non-Hermitian Hamiltonian whose time-evolution operator possesses a dominant Perron--Frobenius eigenmode with equal amplitudes and fixed relative phases across all optical channels~\cite{Perron1907,frobenius1909matrizen,gantmakher1959theory}. 
Under the time evolution of such non-Hermitian Hamiltonian, all other eigencomponents are attenuated relative to this dominant mode, so arbitrary multimode input fields converge to a uniform, phase-locked optical state. 
The convergence rate is determined by the spectral gap between the dominant eigenvalue and the rest of the spectrum, while the total optical throughput can be tuned independently through an overall attenuation factor. 
In this way, engineered dissipation directly performs the work of synchronization.

The paper is organized as follows. 
First, we introduce the class of non-Hermitian transition matrices used for dissipation-induced phase synchronization and show how their spectra determine the synchronized state and convergence rate. 
We then describe how these matrices are embedded into a programmable photonic circuit through unitary dilation. 
Next, we experimentally demonstrate convergence of arbitrary multimode light fields to an equal-intensity, phase-locked collective state. 
Finally, we show independent control of synchronization speed and optical throughput, and discuss how this non-Hermitian approach to synchronizing light can be extended to larger photonic systems and to applications in coherent optical and quantum technologies.

\section*{Theory}

\subsection*{How dissipation drives synchronization}

The mathematical heart of the experiment is a class of matrices called
\emph{row-stochastic}.  A matrix $F$ is row-stochastic if all its entries are non-negative
and each row sums to one:
\begin{equation}
  F_{ij} \ge 0, \qquad \sum_{j} F_{ij} = 1 \quad \text{for every row } i.
  \label{eq:stochastic}
\end{equation}
Such matrices naturally describe how probabilities flow in a Markov chain: the entry $F_{ij}$
is the probability of transitioning from state $i$ to state $j$ in one step.  In our
photonic setting, we apply the same algebraic structure to complex optical field amplitudes
rather than probabilities, giving the framework a non-Hermitian character.

The row-sum condition has an immediate and powerful consequence.  The uniform vector
$v_0 = (1, 1, \dots, 1)^{\mathsf T}$ is always an eigenvector of any row-stochastic matrix,
with eigenvalue $\lambda_0 = 1$.  The Perron--Frobenius
theorem~\cite{gantmakher1959theory} guarantees that for an irreducible, aperiodic matrix
this is the \emph{only} eigenvalue of modulus one; all others satisfy $|\lambda_j| < 1$.
This spectral structure is what makes synchronization inevitable.  Any initial state can be
written as a combination of the eigenvectors of $F$.  After each application of $F$, the
components along eigenvectors with $|\lambda_j| < 1$ are suppressed by a factor of
$|\lambda_j|$, while the component along $v_0$ is unchanged.  Applying $F$ repeatedly
therefore erases all mode-to-mode differences exponentially fast, leaving only the uniform
eigenmode $v_0$.  In physical terms: every output channel ends up with the same complex
amplitude, meaning equal intensities and a single locked phase. The only information from the initial state that survives is the overall complex amplitude
$c_0 = \pi^{\mathsf T} a(0)$, where $\pi$ is the left eigenvector of $F$ associated with
eigenvalue~$1$ (the stationary distribution of the corresponding Markov chain), and $a(0)$
is the initial field vector.  This determines the common amplitude and phase of the
synchronized output but carries no information about how the initial energy was distributed
across modes.  Dissipation has selectively erased the microscopic structure while preserving
the collective global order --- precisely the mechanism that underlies synchronization in
natural systems.

\subsection*{Speed of synchronization and the spectral gap}

How quickly does the system reach the synchronized state?  The answer is influenced by the second
largest eigenvalue modulus, or SLEM, defined as $\sigma = \max_{j \ge 1} |\lambda_j|$.  The
difference $\Delta = 1 - \sigma$ is called the \emph{spectral gap}.  Each application of
$F$ reduces the deviation from synchronization by at least a factor of $\sigma$, so the
characteristic number of steps to reach the synchronized state scales as
\begin{equation}
  k_{\mathrm{sync}} \sim \frac{1}{1 - \sigma} = \frac{1}{\Delta}.
  \label{eq:ksync}
\end{equation}
A large spectral gap means the non-dominant eigenvalues are small, fluctuations are
suppressed quickly, and synchronization is fast.  A small gap means slow convergence.  This
single number, $\Delta$, is the key control parameter.

\subsection*{Continuous-time picture and effective Hamiltonian}

It is natural to reformulate the discrete-step dynamics as a continuous evolution.  We
assign a physical time $t_0$ to each application of the matrix $F$ (one discrete step
corresponds to elapsed time $t_0$), and define an effective non-Hermitian Hamiltonian
\begin{equation}
  H_{\mathrm{eff}} = \frac{\log F}{-i\,t_0},
  \label{eq:Heff}
\end{equation}
where $\log$ is the matrix logarithm.  The state then evolves continuously as
$\psi(t) = e^{-iH_{\mathrm{eff}}\,t}\,\psi(0)$, which exactly reproduces the discrete
iteration $\psi(kt_0) = F^k\psi(0)$ at integer multiples of $t_0$.

The eigenvalues of $H_{\mathrm{eff}}$ have negative imaginary parts for all non-dominant
modes, meaning those modes are exponentially damped.  Only the synchronized eigenmode $v_0$
maps to a zero eigenvalue of $H_{\mathrm{eff}}$ and is therefore preserved indefinitely.
The non-Hermitian character of $H_{\mathrm{eff}}$ is not a nuisance but the precise
mathematical signature of dissipation acting as a synchronizing force.

\subsection*{Independent control of energy and synchronization}

\begin{figure}[ht]
    \centering
    \includegraphics[width=1.0\linewidth]{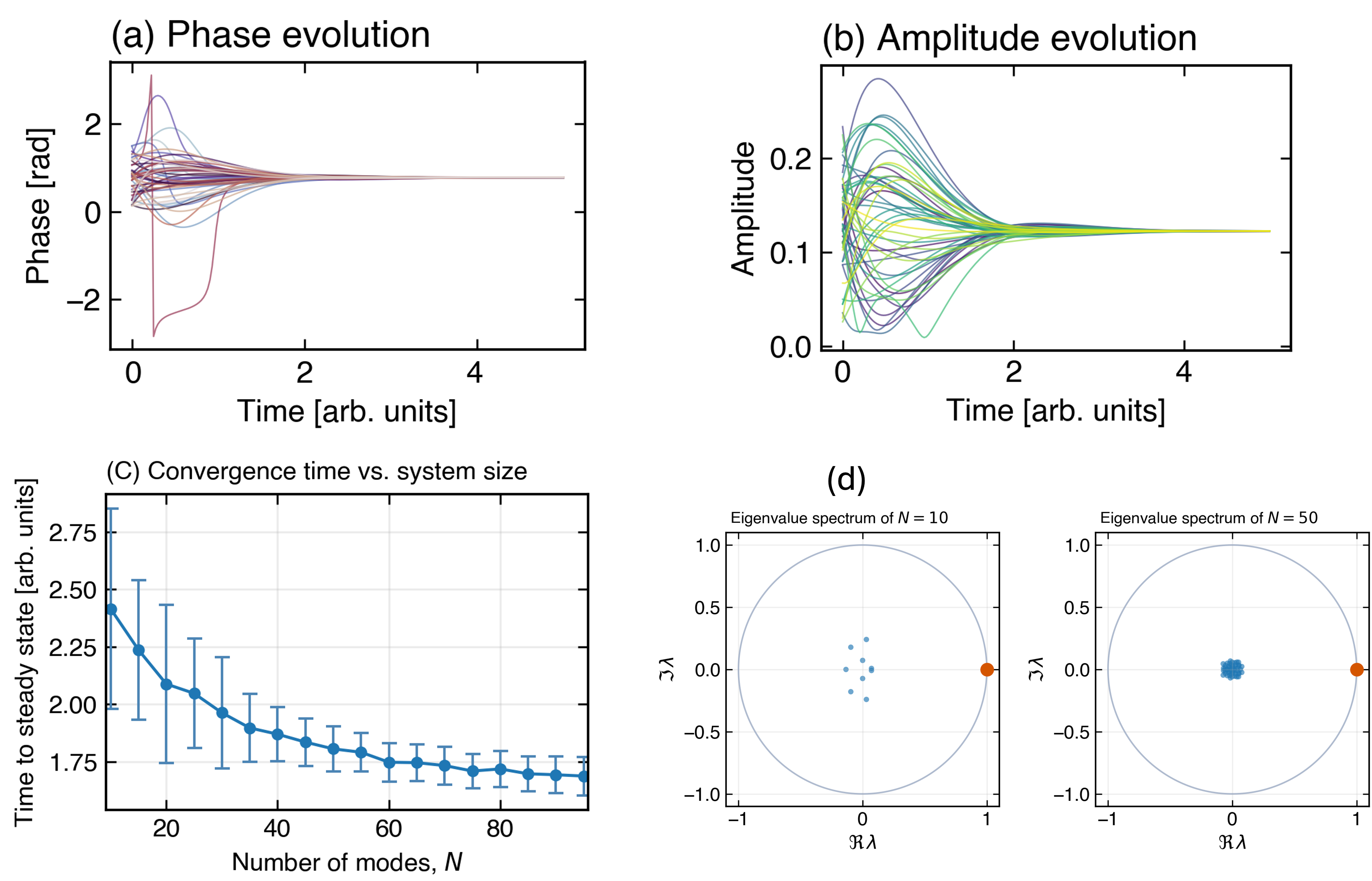}
    \caption{\textbf{Dissipative synchronization through stochastic matrices.}
    \textbf{(a)} Simulated phase evolution for a 50-mode system, showing convergence from a
    random initial distribution to a single global phase.
    \textbf{(b)} Corresponding amplitude evolution, showing equalization of initially
    disordered values.
    \textbf{(c)} Average synchronization time as a function of system size, from 100
    simulations per $N$.  Larger systems synchronize faster because the spectral gap grows
    with dimension: as $N$ increases, the non-dominant eigenvalues are pushed deeper inside
    the unit disk.
    \textbf{(d)} Eigenvalue spectra for $N = 10$ and $N = 50$.  The Perron--Frobenius
    eigenvalue $\lambda_0 = 1$ (red) is isolated; the remaining eigenvalues lie inside the
    unit disk and cluster further from its boundary as $N$ grows.}
    \label{fig:sim}
\end{figure}

Embedding a non-Hermitian map in a passive photonic circuit necessarily attenuates the total
optical power, since dissipation removes energy into the ancilla channels.  However, the
total attenuation and the relative synchronization dynamics can be controlled independently.
Adding a global imaginary offset $\gamma \ge 0$ to the Hamiltonian,
\begin{equation}
  H_{\mathrm{eff}}^{\prime} = H_{\mathrm{eff}} - i\gamma\,I,
  \label{eq:Hprime}
\end{equation}
modifies the evolution to $\psi(t) = e^{-\gamma t}\,e^{-iH_{\mathrm{eff}}\,t}\,\psi(0)$.
The scalar factor $e^{-\gamma t}$ controls the total optical power uniformly across all
modes, while the original operator $e^{-iH_{\mathrm{eff}}\,t}$ continues to govern the
relative amplitudes and phases.  Because the synchronization metric depends only on the
ratios of amplitudes and the differences of phases between modes, not on the global power
level, it is completely unaffected by $\gamma$. Figure~\ref{fig:sim} shows numerical simulations of these dynamics.  Starting from 50 modes
with random initial amplitudes and phases, the system converges to a uniform, phase-locked
state.  Systematically varying $N$ from 10 to 100 reveals a counterintuitive result: larger
systems synchronize faster.  The reason is spectral.  In a row-stochastic matrix of size
$N$, each row must sum to one, which forces the non-dominant eigenvalues to concentrate
closer to the origin as $N$ grows, widening the spectral gap and accelerating convergence.

\subsection*{Programmable synchronization speed}

To gain deterministic control over synchronization speed, we engineer a structured
one-parameter family of matrices:
\begin{equation}
  F(\alpha) = \alpha\,I + (1-\alpha)\,J, \qquad J_{ij} = \tfrac{1}{N},
  \label{eq:family}
\end{equation}
where $I$ is the identity and $J$ is the uniform matrix.  By construction the SLEM is
exactly $\alpha$, so tuning this single parameter continuously adjusts the spectral gap
$\Delta = 1 - \alpha$ while leaving the target synchronized eigenmode $v_0$ unchanged.
All non-dominant fluctuations decay as $\alpha^k$ per step, giving a threshold
synchronization time
\begin{equation}
  t_{\mathrm{thres}} = \frac{t_0}{|\ln\alpha|}\,
  \ln\frac{A}{\varepsilon},
  \label{eq:tthres}
\end{equation}
where $\varepsilon$ is the convergence threshold and $A$ is a prefactor that depends
on the overlap between the input state and the Perron eigenvector.  The dominant
factor $1/|\ln\alpha| \approx 1/\Delta$ is controlled entirely by the SLEM; the
initial-state dependence enters only through a logarithmic correction in the numerator.
Because $F(\alpha)$ is symmetric, its eigenvectors are orthogonal and the SLEM alone
determines the decay rate --- this clean single-parameter control is what our experiment
exploits.

In general, however, the synchronization time is not governed by the spectral gap alone.
For a generic (non-symmetric) row-stochastic matrix with SLEM $\sigma$, the threshold time
obeys the upper bound
\begin{equation}
  t_{\mathrm{thres}} \lesssim \frac{t_0}{|\ln\sigma|}\,
  \ln\!\left(\frac{C_F\,\|\psi_0\|_2}
  {\varepsilon\,|\pi^{\mathsf T}\psi_0|}\right),
  \label{eq:tthres_general}
\end{equation}
where $C_F$ is a matrix-dependent constant bounded by the condition number of the
eigenvector matrix and defined in the \textbf{Supplementary information}, encoding the degree of non-normality of $F$, and
$\pi^{\mathsf T}\psi_0$ is the projection of the input onto the stationary distribution.
Two matrices sharing the same SLEM can synchronize at very different rates if their
eigenvector geometries differ --- an effect analyzed quantitatively in the \textbf{Supplementary Information}.

\section*{Physical implementation in photonic meshes}

\subsection*{Embedding dissipation in a unitary network}

A non-Hermitian transformation cannot be implemented directly in a passive linear optical
circuit, because every physical beamsplitter network is unitary.  The standard solution is
to embed the non-unitary matrix $T$ inside a larger unitary that acts on both the system
modes and a set of auxiliary (ancilla) modes initialized in vacuum~\cite{tischler2018quantum}.
Tracing out the ancilla modes after the transformation reproduces the intended non-unitary
action on the system alone. The embedding works via the singular value decomposition $T = U\Sigma W^\dagger$.  The
unitary factors $U$ and $W$ are implemented directly as interferometer meshes.  Each
singular value $\sigma_j \le 1$ is realized by coupling the corresponding system mode to a
vacuum ancilla on a beamsplitter of transmissivity $\tau_j = \sigma_j^2$: the ancilla
absorbs the energy that would otherwise violate unitarity, and the system mode emerges
attenuated by exactly $\sigma_j$.  The complete circuit --- input interferometer, ancilla
couplings, output interferometer --- is globally unitary; the dissipation appears only in the
effective description of the system modes after the ancillas are
discarded~\cite{xu2025non}.

\begin{figure}[htbp]
    \centering
    \includegraphics[width=1.0\linewidth]{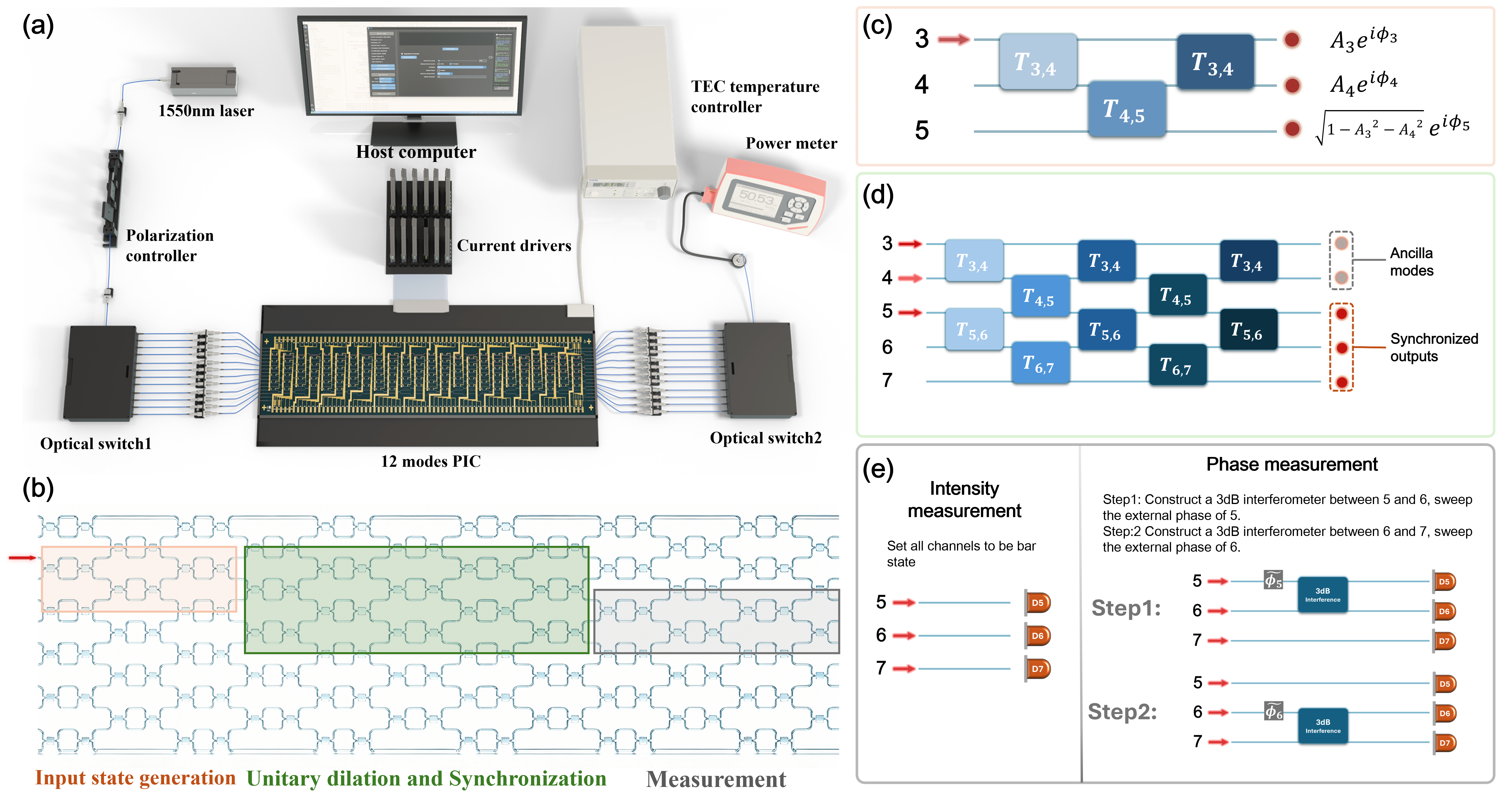}
    \caption{\textbf{Experimental platform for non-Hermitian synchronization.}
    \textbf{(a)} Automated experimental setup: a 1550\,nm pulsed laser excites a $12\times
    12$ silicon MZI mesh via MEMS switches; output intensities are read sequentially with a
    single power meter.
    \textbf{(b)} Functional layout of the chip, partitioned into input state preparation
    (orange), synchronization via unitary dilation (green), and reconfigurable measurement
    (gray).
    \textbf{(c)} Input module: three MZIs distribute one laser input across three waveguide
    modes with independently programmable amplitudes and phases.
    \textbf{(d)} Synchronization module: a $3\times 3$ non-Hermitian operator is embedded as
    a block in a $5\times 5$ unitary network; ancilla outputs at ports 3 and 4 carry the
    dissipated power to non-measured ports.
    \textbf{(e)} Measurement module: configured either for direct intensity detection (left)
    or for phase extraction via interference sweeps in a 3-dB MZI (right).}
    \label{fig:2}
\end{figure}

\subsection*{Silicon photonic platform}

The experiments are performed on a $12\times 12$ reconfigurable interferometer mesh
fabricated in a CMOS-compatible silicon photonics process (Advanced Micro Foundry),
Fig.~\ref{fig:2}(a).  The chip integrates 66 thermally tunable Mach--Zehnder interferometers
(MZIs), each comprising two multimode-interference couplers linked by a phase-controlled
arm, making every MZI a universal $2\times 2$ optical element.  A pulsed diode laser at
1550\,nm injects light into selected input ports via a $1\times 12$ MEMS switch, and a
second MEMS switch sequentially connects output ports to a calibrated power meter (Thorlabs
PM400).  Phase control of each MZI is provided by thermal phase shifters driven by
programmable electronic drivers (Qontrol, $20\,\mu$A resolution). As illustrated in Fig.~\ref{fig:2}(b), the 12-mode chip is organized into three functional
zones.  The \textbf{input module} (orange) uses three MZIs to distribute light from a single
input port across three waveguide modes with arbitrary amplitudes and phases,
Fig.~\ref{fig:2}(c).  The \textbf{synchronization module} (green) embeds the target $3\times
3$ non-Hermitian operator within a $5\times 5$ unitary network spanning ten MZIs: the three
signal modes enter at rows 3--5, the synchronized outputs emerge at rows 5--7, and the two
ancilla outputs at rows 3--4 carry the dissipated flux to unmeasured ports,
Fig.~\ref{fig:2}(d).  The \textbf{measurement module} (gray) can be reconfigured on-the-fly
between direct intensity detection and high-precision phase characterization,
Fig.~\ref{fig:2}(e).

\section*{Results}

\subsection*{Amplitude and phase synchronization}

\begin{figure}[ht]
    \centering
    \includegraphics[width=1.0\linewidth]{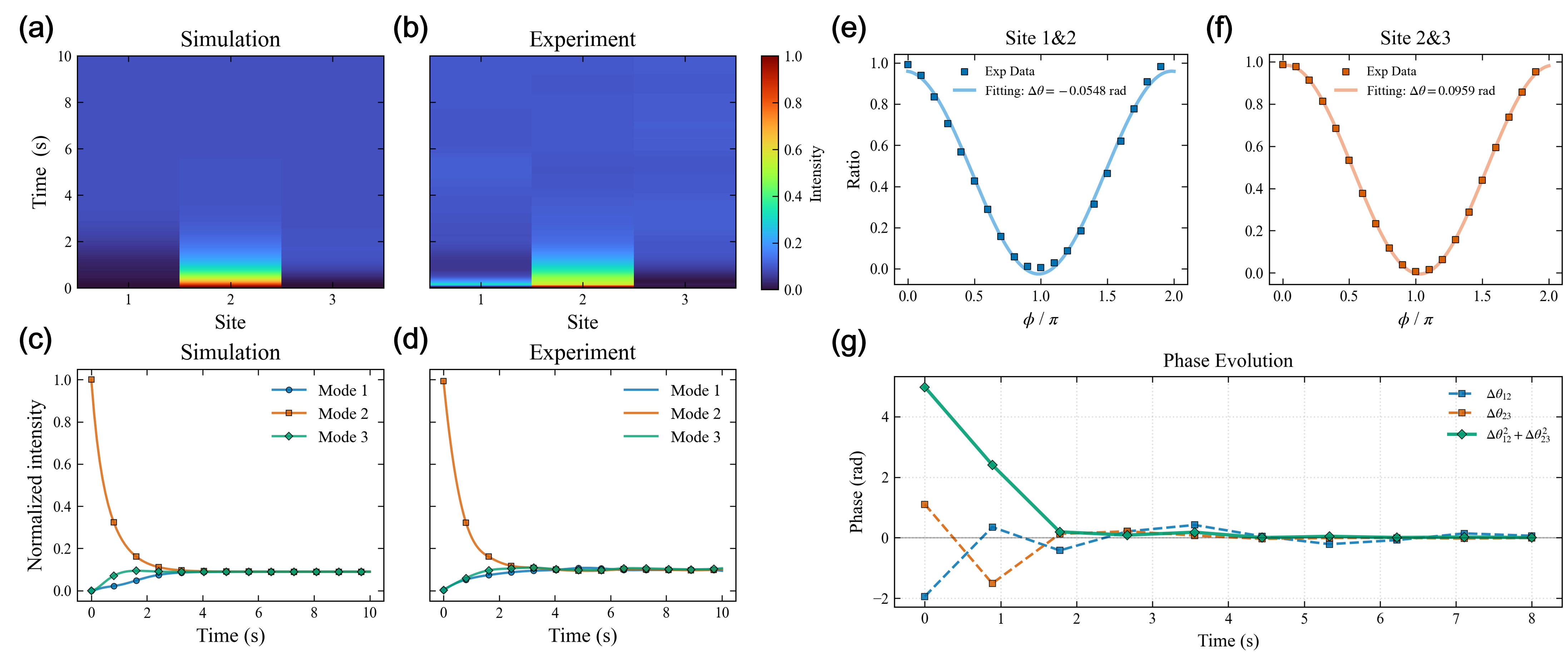}
    \caption{\textbf{Experimental demonstration of amplitude and phase synchronization.}
    \textbf{(a,b)} Simulated (a) and measured (b) intensity evolution across three modes,
    initialized with all power in Mode~2.  The engineered dissipation redistributes the
    intensities to a uniform steady state.
    \textbf{(c,d)} Normalized intensity trajectories confirm quantitative agreement between
    theory and experiment.
    \textbf{(e,f)} Phase characterization at the synchronized steady state: interference
    fringes (solid lines through data points) from the reconfigurable phase module yield the
    relative phase differences $\Delta\theta_{12}$ and $\Delta\theta_{23}$.
    \textbf{(g)} Temporal evolution of the phase differences and the combined variance metric
    $\Delta\theta_{12}^2 + \Delta\theta_{23}^2$, which collapses to zero as synchronization
    is reached.}
    \label{fig:3}
\end{figure}

We first verify the basic synchronization effect.  All optical power is loaded into a single
mode (Mode~2), and the stochastic matrix is applied repeatedly.  As shown in
Fig.~\ref{fig:3}(a,b), the intensity spreads progressively across all three modes, reaching
a uniform distribution in excellent agreement with simulation.  The normalized trajectories
in Fig.~\ref{fig:3}(c,d) confirm quantitative agreement. Phase synchronization is verified using the reconfigurable measurement module as a
two-channel interferometer.  We apply a controllable phase shift $\phi$ to one arm of a 3-dB
MZI and record the output intensity as $\phi$ is swept from $0$ to $2\pi$ in 20 steps.  The
resulting fringe
\begin{equation}
  I(\phi) = \tfrac{1}{2}\bigl(A_1^2 + A_2^2 + 2A_1 A_2 \cos(\phi - \Delta\theta)\bigr)
  \label{eq:fringe}
\end{equation}
is fitted numerically to extract the relative phase $\Delta\theta$ between the two selected
modes, even in the presence of residual amplitude imbalance or stray light.  The procedure is
applied independently to each adjacent pair (modes 1\&2 and modes 2\&3), yielding
$\Delta\theta_{12}$ and $\Delta\theta_{23}$. The results in Fig.~\ref{fig:3}(e--g) are clear.  At the final step, both phase differences
are negligible, confirming global phase locking.  Tracking the combined phase variance
$\Delta\theta_{12}^2 + \Delta\theta_{23}^2$ over time shows a rapid collapse to zero,
providing a quantitative signature of synchronization.

\subsection*{Programmable synchronization speed}

\begin{figure}[htbp]
\centering
\includegraphics[width=1.0\linewidth]{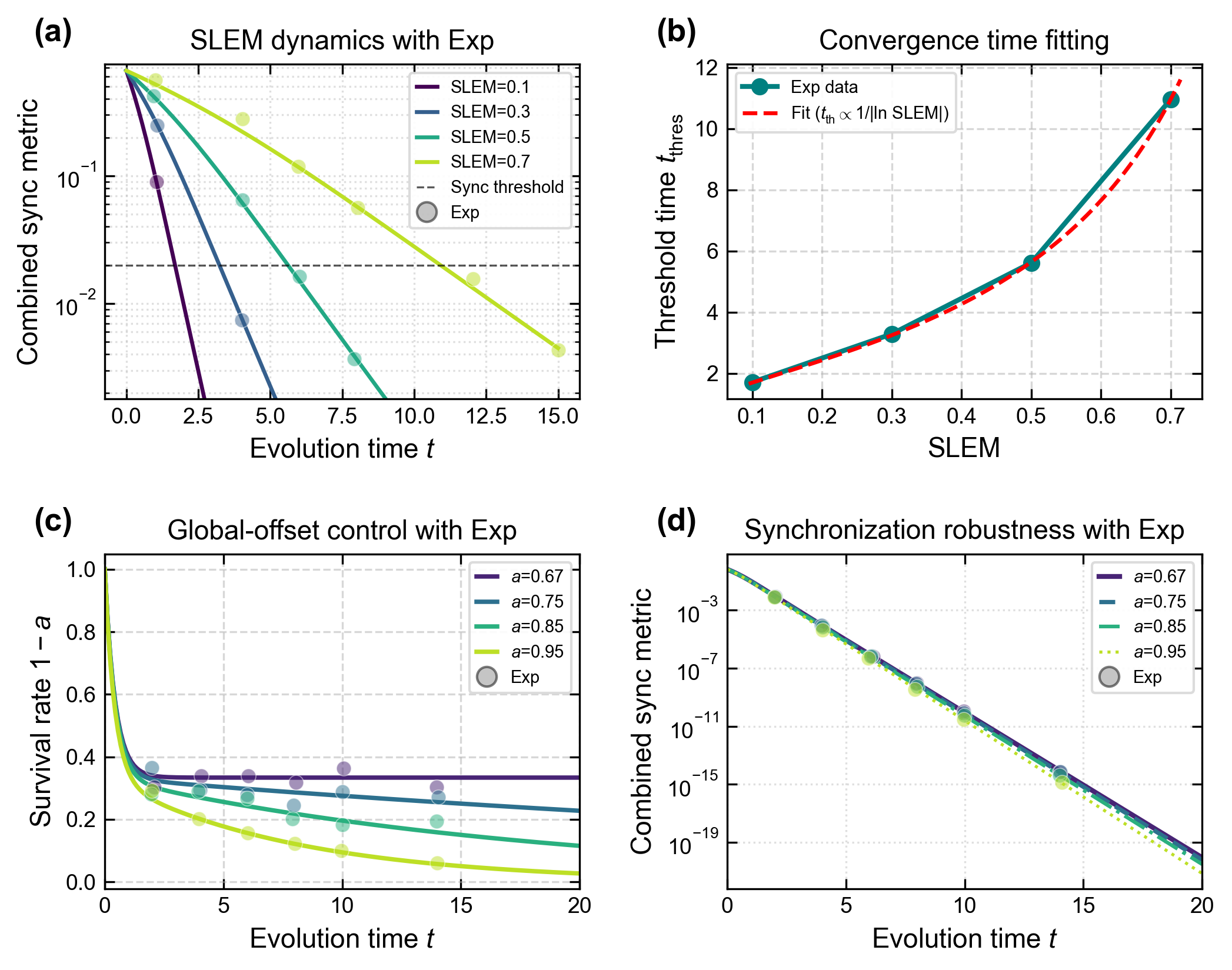}
\caption{\textbf{Programmable synchronization velocity and decoupled energy control.}
\textbf{(a)} Synchronization metric versus step number for different SLEM values $\alpha$.
Experimental data (circles) follow the theoretical exponential decay envelopes (lines),
with faster decay for smaller $\alpha$ (larger spectral gap).
\textbf{(b)} Threshold time $t_{\mathrm{thres}}$ versus SLEM.  The measured values confirm
the scaling $t_{\mathrm{thres}} \propto 1/|\ln\alpha|$ across all tested configurations.
Deviations expected for generic non-symmetric matrices (due to structural amplification
effects) are discussed in the \textbf{Supplementary Information}.
\textbf{(c)} Measured energy survival rate for different prescribed offsets $\gamma$.  The
asymptotic power level tracks the target exactly.
\textbf{(d)} Synchronization metric for the same runs as (c).  All curves collapse onto one
master trajectory regardless of the energy loss, confirming that power and phase locking are
independently controlled.}
\label{fig:4}
\end{figure}

Having confirmed synchronization, we now show that its speed is fully programmable.  Using
the one-parameter family $F(\alpha)$ from Eq.~\eqref{eq:family}, we set different values of
the SLEM $\alpha$ while keeping the target eigenmode fixed.  Figure~\ref{fig:4}(a) shows the
synchronization metric decaying over time for several values of $\alpha$: smaller $\alpha$
means a larger spectral gap and faster convergence, in close agreement with the theoretical
exponential envelopes.  Figure~\ref{fig:4}(b) plots the extracted threshold times and
confirms the predicted inverse scaling $t_{\mathrm{thres}} \propto 1/|\ln\alpha|$. 

Because $F(\alpha)$ is symmetric, the SLEM is the sole rate-controlling parameter and our
experiment cleanly isolates its role.  For generic non-symmetric matrices, additional
factors --- notably the eigenvector geometry encoded by $C_F$ in
Eq.~\eqref{eq:tthres_general} --- can delay synchronization beyond the ideal spectral-gap
prediction; this richer behaviour is analyzed in the \textbf{Supplementary Information}. Finally, we demonstrate decoupled energy control.  We implement the modified Hamiltonian
$H_{\mathrm{eff}}^{\prime} = H_{\mathrm{eff}} - i\gamma\,I$ for several values of $\gamma$.
Figure~\ref{fig:4}(c) shows that the total power decays to a prescribed target level in each
case.  Figure~\ref{fig:4}(d) shows the synchronization metric for the same runs: all curves
collapse exactly onto a single master trajectory, regardless of how aggressively the total
power is attenuated.  This confirms the predicted decoupling between energy loss and coherent
phase-locking dynamics.

\section*{Discussion}

We have demonstrated that programmable photonic meshes can engineer dissipation-induced phase
synchronization on chip, with full control over both the synchronization dynamics and the
energy throughput.  The core result is conceptually simple: by implementing transition
matrices whose spectral structure is governed by the Perron--Frobenius theorem, we ensure
that any input light field is driven to a unique, phase-locked collective state.  The rate of
convergence is set by a single spectral parameter --- the gap between the dominant and
subdominant eigenvalues --- and can be dialed continuously, from slow to fast, by adjusting
the matrix design. From a fundamental standpoint, this work establishes a clean linear-algebraic framework for
synchronization phenomena that are typically studied through nonlinear dynamical models.
In our setting the mechanism is transparent: dissipation systematically suppresses all
eigenmodes of the transition matrix except the one that is protected by the row-sum symmetry,
and that protected mode is precisely a uniform, phase-coherent state.  Dissipation is not a
limitation of the system; it is the mechanism that enforces order. The platform has practical relevance as well.  Maintaining stable phase relationships across
multiple optical channels is essential in dense wavelength-division multiplexing,
co-packaged optics, and quantum photonic networks, and our approach offers a route to
achieving this in a chip-scale, fully reconfigurable format.  Interestingly, the spectral
analysis predicts that synchronization naturally becomes more efficient as the system grows:
larger mode-density configurations have wider spectral gaps, meaning they synchronize faster
for free. Scaling to larger systems does introduce practical challenges.  Chief among them is thermal
crosstalk: heat from one phase shifter diffuses into adjacent elements, inducing parasitic
phase shifts that can corrupt the engineered stochastic matrix.  The collective nature of the
Perron--Frobenius dynamics provides some inherent robustness --- small perturbations to
individual matrix entries do not shift the synchronized eigenmode significantly --- but
large-scale implementations will likely require active thermal compensation. More broadly, this work joins a growing body of research showing that dissipation, handled
carefully, is a powerful tool rather than an obstacle.  Reservoir engineering in
superconducting circuits, loss-induced state preparation in quantum optics, and now
programmable synchronization in photonic meshes all share the same underlying logic.  We
hope the clear experimental framework and the transparent connection to the Perron--Frobenius
spectral theory will make this an accessible entry point for exploring dissipative collective
phenomena in a broad range of engineered physical systems.

\subsection*{Acknowledgments}

Z.X., A.C., G.K., and A.W.E.\ acknowledge support from the Knut and Alice Wallenberg (KAW)
Foundation through the Wallenberg Centre for Quantum Technology (WACQT).
J.G.\ acknowledges support from the Swedish Research Council (Ref.\ 2023-06671 and
2023-05288), Vinnova (Ref.\ 2024-00466), and the Göran Gustafsson Foundation.
A.W.E.\ acknowledges support from the Swedish Research Council (VR) Starting Grant
(Ref.\ 2016-03905).

\subsection*{Competing Interests}
The authors declare no competing interests.

\subsection*{Author Contributions}
A.W.E.\ and Z.X.\ wrote the paper, with input from all other authors.

\end{document}